\documentstyle[preprint,aps,psfig]{revtex}

\begin{document}

\tighten

\preprint{NYU-TH/00/03/08}

\title{SO(10) Theory of R-parity and Neutrino Mass}

\author{
Charanjit S. Aulakh$^{(1)}$, Borut Bajc$^{(2,3)}$, 
Alejandra Melfo$^{(4)}$, Andrija Ra\v{s}in$^{(5)}$ and
Goran Senjanovi\'c$^{(6,2)}$}
\address{$^{(1)}$ {\it Dept. of Physics, Panjab University,
Chandigarh, India   }}
\address{$^{(2)}$ {\it Department of Physics, New York University, New 
York, NY 10003, U.S.A. }}
\address{$^{(3)}$ {\it J. Stefan Institute, 1001 Ljubljana, Slovenia}}
\address{$^{(4)}$ {\it CAT, Facultad de Ciencias, Universidad de
Los Andes, M\'erida, Venezuela }}
\address{$^{(5)}${\it Dept. of Physics and Astronomy, University of
North Carolina, Chapel Hill, NC 27599, U.S.A. }}
\address{$^{(6)}${\it International Center for Theoretical Physics,
Trieste, Italy }}

\maketitle
\begin{abstract}
We study the Higgs sector of a SO(10) grand unified theory which 
predicts exact conservation of R-parity at all scales and 
incorporates the see-saw mechanism. We find possible intermediate 
scales and light states compatible with the constraints coming 
from the running of the gauge couplings. Such a pattern could 
lower the SO(10) breaking scale, allowing the $d=6$ proton decay 
operators to be comparable in magnitude to the $d=5$ ones. 

\today
\end{abstract} 

\vspace{0.3cm}

\section {  Introduction}

It is well-known that SO(10) grand unified theory offers an 
appealing framework for the unification of quarks and 
leptons and their forces.  It is the minimal theory which 
unifies a family of fermions in a simple irreducible spinorial
representation. The left-right (LR) symmetry in the form of 
charge conjugation \cite{dva} is a finite gauge  
transformation and thus automatically built into the theory. Last but
not least, it naturally incorporates the see-saw mechanism \cite{seesaw}
which provides a rationale for the smallness of the neutrino mass.

However, in its non-supersymmetric realization it fails to 
provide a ``canonical'' see-saw formula \cite{ms81}. It is 
here that supersymmetry (for a recent review on supersymmetric 
grand unified theories see for example \cite{mohapatra99})
plays an important role and in certain cases the see-saw can take the 
canonical form. We study this issue in the context of the SO(10)
theory and find, unfortunately, a situation similar to the ordinary
case. More important, the see-saw mechanism actually 
gives us the low energy effective theory of supersymmetry. 
What happens is that R-parity remains an exact symmetry at all
scales. This is a fundamental result which guarantees the 
stability of the lightest supersymmetric partner (LSP), an ideal 
candidate for the dark matter of the universe. This is true of any 
renormalizable theory of the see-saw mechanism based on the 
spontaneous breaking of B$-$L symmetry \cite{amrs99}.

Without unification one can not predict the value of 
the right-handed neutrino mass scale. In this sense  SO(10) is an 
ideal theory especially in its supersymmetric version. It 
incorporates all the above features and it helps narrow down 
the range in which the right-handed neutrino mass lies. 

In this paper we construct a complete  SO(10) theory of see-saw 
mechanism and R-parity. After a careful study of symmetry-breaking 
patterns, we compute the particle spectra and perform the 
unification analysis. We find a plethora of states (often 
carrying color or a large electromagnetic charge) whose 
masses, due to supersymmetry, could in principle lie much 
below the associated symmetry breaking scales. The reason for this 
is the violation of the survival principle, which we discussed 
at length in \cite{abmrs99}. There we have coined the term  
``survival of the fittest'' for this phenomenon. It is due to the
absence of some quartic couplings in the potential (i.e., the absence
of some trilinear couplings in the superpotential), as is often the
case in supersymmetric models. 

The rest of this paper is organized as follows: in the next section 
we discuss the salient features of the  SO(10) theory with a  
renormalizable see-saw and the possible symmetry breaking 
patterns. In Section III we give detailed analysis of the 
particle spectrum and in Section IV we present the unification 
constraints. In Section V the phenomenological and cosmological 
consequences are discussed at some lengths. Section VI is devoted to 
the summary. The technical and computational details 
of symmetry breaking are left for the Appendices.

\section{ The SO(10) supersymmetric theory with the 
renormalizable see-saw mechanism.} 
   
Supersymmetric SO(10) models have been studied at length 
\cite{so10}, but almost exclusively with the non-renormalizable 
version of the see-saw mechanism. More precisely, one chooses 
one (or more) pair of Higgs fields in the spinorial representation 
{\bf 16} and $\overline{\bf 16}$ whose VEVs induce B-L breaking 
and the mass for the right-handed neutrino through the $d=4$ 
terms: $m_{\nu_R} \simeq \langle{\bf 16} \overline{\bf 16}
\rangle /M_{Pl}$. The disadvantage of this program is that 
then R-parity is broken at a large scale $M_R$, and thus one 
needs additional, often ad-hoc symmetries to understand the 
smallness of R-parity breaking at low energies.

Our motivation is orthogonal to this. We wish to have a theory 
of R-parity and this points immediately to the renormalizable 
version of the see-saw mechanism. To see this, recall first 
that under R-parity  $p\to p$, $\tilde p \to - \tilde p$ (where 
$p$ stands for particles and $\tilde p$ for sparticles of the
supersymmetric standard model) can be written as ($S$ is the spin)

\begin{equation}
R = (-1)^{3(B-L) + 2 S} \equiv M (-1)^{2S}\;,
\label{rparity}
\end{equation}
where the so-called M parity is obviously equivalent to R 
(the factor $(-1)^{2S}$ becomes 1 for the physical Hamiltonians 
and only scalars with $S=0$ are allowed to have non-vanishing VEVs).

One of the most appealing aspects of the supersymmetric  SO(10) is that 
M is a finite gauge symmetry, since under M
\begin{eqnarray}
{\bf 16} \; \stackrel{M}{\longrightarrow} \;  -{\bf 16} \quad , \quad
{\bf 10} \; \stackrel{M}{\longrightarrow} \;  {\bf 10}\;,
\label{mparity}
\end{eqnarray}

\noindent 
and all other representations built out of the fundamental
{\bf 10}, 
such as {\bf 45}, {\bf 54}, {\bf 126}, etc. are even. The symmetry
in (\ref{mparity})  
is simply $C^2$, where $C$ is the center of  SO(10), so that under
it {\bf 16}$ \to i$ {\bf 16}, {\bf 10}$ \to -${\bf 10}. 
This points strongly towards using a 126-dimensional Higgs for the
 breaking of B$-$L and the see-saw mechanism
(for previous analysis of SO(10) models with {\bf 126} 
representations see \cite{aulamoha,leem94,sato}). One 
drawback of these representations is their huge contribution 
to the SO(10) $\beta$ function, so that the Landau pole is 
reached very soon above $M_X$. We will be interested only 
in the physics below or at the unification scale, where the 
couplings are perturbative. 

We wish to construct a renormalizable  SO(10) theory with a 
see-saw, and this requires the minimum set of Higgs 
representations which break  SO(10) down to the MSSM:

\begin{equation}
S = {\bf 54} \;, \quad A = {\bf 45} \; , \quad \Sigma = {\bf 126} \; , \quad 
\overline\Sigma= \overline{\bf 126} \;.
\label{higgses}
\end{equation}

Although  SO(10) is anomaly-free, as is well-known, one has to use both 
$\Sigma$  and $\overline\Sigma$ in order to ensure the flatness  of the
D-piece
of the potential at large scales $\gg M_W$.

What are the possible channels of symmetry breaking and why do we need  all
the above fields? Of course, one possible channel strongly encouraged 
by the MSSM unification constraints is through SU(5). Another possibility,
 much less studied, is the intermediate Pati-Salam (PS) and/or LR scale.
For this purpose, it
is useful to know the decomposition of the above fields under the
PS gauge group SU(2)$_L\times$ SU(2)$_R \times$ SU(4)$_C$

\begin{eqnarray}
S & = {\bf 54} &= (1,1,1) + (1,1,20) +(3,3,1) + (2,2,6)\;, \nonumber \\
A & = {\bf 45}  & = (1,1,15) + (3,1,1) + (1,3,1) + (2,2,6)\;,  \nonumber \\ 
\Sigma  & = {\bf 126} & = (3,1,\overline{10}) + (1,3,10) + (2,2,15) +
(1,1,6)\;,\nonumber \\
\overline\Sigma  & = \overline{\bf 126} & = (3,1, 10) + 
(1,3,\overline{10}) +(2,2,15) +(1,1,6)\;.
\end{eqnarray}

One may think that it is redundant to take both $A$ and $S$ 
together with $\Sigma$ and $\overline \Sigma$. After all either 
$(1,1,1)$ or $(1,1,15)$ could produce the first stage of symmetry
breaking, down to the PS or LR symmetry, to be followed by 
$(1,3,10)$ in $\Sigma$. Notice that the $(1,3, 10)$ field contains 
a color singlet component $\Delta_c$ which couples to the 
right-handed neutrino $\nu^c$ in {\bf 16}, and through 
$\langle \Delta_c \rangle \equiv M_R$ produces the see-saw 
mechanism. Since $\nu^c$ is an SU(5) singlet, so is obviously 
$\langle \Delta_c\rangle$. Thus the possible $\langle {\bf 126} 
\rangle \equiv \langle \Delta_c\rangle \neq 0$ can break SO(10) 
only down to SU(5).

Clearly, either $\langle A \rangle \neq 0 \neq \langle \Sigma \rangle$
or  $\langle S \rangle \neq 0 \neq \langle \Sigma \rangle$ suffices
 to break  the  SO(10) symmetry all the way down to the MSSM. So why
is it that we need  both $A$ and $S$ fields in the renormalizable
version of the theory which we study here ? 

To see this, we need to write down the most general superpotential
containing $S, A, \Sigma $ and $\overline\Sigma$

\begin{eqnarray}
W &=&{m_S\over 2}{\rm Tr}\, S^2+{\lambda_S \over 3} {\rm Tr}\, S^3
+{m_A\over 2}{\rm Tr}\, A^2+\lambda{\rm Tr}\, A^2 S \nonumber \\
& &+m_\Sigma\Sigma\overline\Sigma +\eta_S\Sigma^2 S + 
\overline\eta_S {\overline\Sigma}^2 S+
\eta_A \Sigma\overline\Sigma A\;.
\label{superpot}
\end{eqnarray}

\noindent The above form is only symbolic, details are given in 
Appendix A.

It is trivial to see why $A, \Sigma $ and $\overline\Sigma$ cannot
suffice. From 
\begin{equation}
F_A = m_A A + \eta_A \Sigma \overline\Sigma
\label{efea}
\end{equation}
and the fact that $\langle \Sigma\overline\Sigma \rangle$ 
preserves SU(5), it is clear that $ \langle A \rangle$ must be 
SU(5) invariant, too. In order to achieve more breaking, one may 
resort to  non-renormalizable operators of the type ${\rm Tr}\, A^4/M$, 
${\rm Tr}\,(\Sigma\overline\Sigma)^2/M$, etc. (where $M$ is some 
new large scale). We prefer to include another field $S$; after all 
if $\langle S \rangle \gg\langle A \rangle$, we will get the 
non-renormalizable terms after integrating out the heavy fields. 

It is also straightforward to see why $S$ does not suffice together 
with $\Sigma$ and $\overline\Sigma$. The point is that the $\eta_S$ 
and $\overline\eta_S$ terms cannot produce any interaction between 
the VEV acquiring fields $(1,1,1)_S$ and $(1,3, 10)_\Sigma$: there 
is no  singlet in the tensor product $10 \times 10$ in SU(4); the 
same argument applies for $\overline\Sigma$. Thus, although 
$\langle S\rangle$ can break  SO(10) down to PS, the lack of 
interactions for the $(1,3,10)$ and $(1,\overline{3},10)$ fields 
forces their VEVs to vanish.

In short, in the renormalizable theory one needs all the above 
fields. One can envision the following physically interesting 
patterns of symmetry breaking, which we label {\bf (a)} and {\bf (b)}
 
\setlength{\unitlength}{0.00083300in}%
\begingroup\makeatletter\ifx\SetFigFont\undefined
\def\x#1#2#3#4#5#6#7\relax{\def\x{#1#2#3#4#5#6}}%
\expandafter\x\fmtname xxxxxx\relax \def\y{splain}%
\ifx\x\y   
\gdef\SetFigFont#1#2#3{%
  \ifnum #1<17\tiny\else \ifnum #1<20\small\else
  \ifnum #1<24\normalsize\else \ifnum #1<29\large\else
  \ifnum #1<34\Large\else \ifnum #1<41\LARGE\else
     \huge\fi\fi\fi\fi\fi\fi
  \csname #3\endcsname}%
\else
\gdef\SetFigFont#1#2#3{\begingroup
  \count@#1\relax \ifnum 25<\count@\count@25\fi
  \def\x{\endgroup\@setsize\SetFigFont{#2pt}}%
  \expandafter\x
    \csname \romannumeral\the\count@ pt\expandafter\endcsname
    \csname @\romannumeral\the\count@ pt\endcsname
  \csname #3\endcsname}%
\fi
\fi\endgroup
\begin{equation}
\begin{picture}(4895,4691)(151,-4317)
\thicklines
\put(3908,-1774){\makebox(6.6667,10.0000){\SetFigFont{10}{12}{rm}.}}
\put(2883,-3482){\vector( 0,-1){569}}
\put(2883, 47){\vector( 0,-1){569}}
\put(2713,-863){\vector(-1,-1){854}}
\put(3168,-863){\vector( 1,-1){854}}
\put(1916,-2173){\vector( 1,-1){854}}
\put(3965,-2173){\vector(-1,-1){854}}
\put(265,-2913){\makebox(0,0)[lb]{\smash{\SetFigFont{14}{16.8}{rm}(a)}}}
\put(5046,-2913){\makebox(0,0)[lb]{\smash{\SetFigFont{14}{16.8}{rm}(b)}}}
\put(2542,218){\makebox(0,0)[lb]{\smash{\SetFigFont{12}{16.8}{rm}SO(10)}}}
\put(1859,-1319){\makebox(0,0)[lb]{\smash{\SetFigFont{11}{13.2}{rm}$M_C$}}}
\put(3737,-1319){\makebox(0,0)[lb]{\smash{\SetFigFont{11}{13.2}{rm}$M_R$}}}
\put(3111,-3767){\makebox(0,0)[lb]{\smash{\SetFigFont{11}{13.2}{rm}$M_W$}}}
\put(2200,-4279){\makebox(0,0)[lb]{\smash{\SetFigFont{12}{16.8}{rm} 
 SU(3)$_C  \times $ U(1)$_{em}$}}}
\put(1973,-3368){\makebox(0,0)[lb]{\smash{\SetFigFont{12}{16.8}{rm}SU(2)$_L
\times  $U(1)$_Y  \times  $ SU(3)$_C$}}} 
\put(3453,-2059){\makebox(0,0)[lb]{\smash{\SetFigFont{12}{16.8}{rm}SU(2)$_L
\times  $ U(1)$_R  \times$  SU(4)$_C$}}} 
\put(-51,-2059){\makebox(0,0)[lb]{\smash{\SetFigFont{12}{16.8}{rm}SU(2)$_L
\times$ SU(2)$_R \times$ SU(3)$_C \times$ U(1)$_{BL}$}}} 
\put(1859,-750){\makebox(0,0)[lb]{\smash{\SetFigFont{12}{16.8}{rm}SU(2)$_L
\times$  SU(2)$_R  \times$  SU(4)$_C$}}} 
\put(2428,-237){\makebox(0,0)[lb]{\smash{\SetFigFont{11}{13.2}{rm}$\langle
S\rangle$}}} 
\put(3111,-237){\makebox(0,0)[lb]{\smash{\SetFigFont{11}{13.2}{rm}$M_X$}}}
\put(2769,-1319){\makebox(0,0)[lb]{\smash{\SetFigFont{11}{13.2}{rm}$\langle
A\rangle$}}} 
\put(2585,-2685){\makebox(0,0)[lb]{\smash{\SetFigFont{11}{13.2}
{rm}$\langle\Sigma\rangle$}}} 

\put(2897,-2685){\makebox(0,0)[lb]{\smash{\SetFigFont{11}{13.2}{rm},
$\langle\overline{\Sigma}\rangle$}}} 
\put(3851,-2685){\makebox(0,0)[lb]{\smash{\SetFigFont{11}{13.2}{rm}$M_C$}}}
\put(1802,-2685){\makebox(0,0)[lb]{\smash{\SetFigFont{11}{13.2}{rm}$M_R$}}}
\put(2428,-3767){\makebox(0,0)[lb]{\smash{\SetFigFont{11}{13.2}
{rm}$\langle\Phi\rangle$}}} 
\end{picture}
\label{chain}
\end{equation}

It turns out that both of these chains are quite interesting in the
sense that they lead to a plethora of ``light'' states, i.e., 
states whose masses lie below the corresponding symmetry breaking scale.

In order to study any of the chains, we need the conditions for the F
and D  
flatness at the scales $\gg M_W$. This again is discussed at length in 
Appendix A, here we address only the salient features. The F-flatness
equations are

\begin{eqnarray}
F_{(1,1,1)_S} = m_s s + {1 \over 2} \lambda_S s^2 + {2\over 5}
 \lambda
 (a^2 -b^2) &=&0 \;,\nonumber \\
& & \nonumber \\
F_{(1,1,15)_A} = m_A a + 2 \lambda a s + {1 \over 2} \eta_A
\sigma\bar\sigma 
&=& 0 \;,\nonumber \\
& &  \nonumber \\
F_{(1,3,1)_A} = m_A b - 3 \lambda b  s + {1\over 2} \eta_A \sigma
\bar\sigma  
&=& 0 \;,\nonumber \\
& & \nonumber \\
F_{(1,3,10)_\Sigma} = \sigma \left[ m_\Sigma + \eta_A (3 a + 2 b) \right]
&=& 0 \;,\nonumber \\
&  & \nonumber \\ 
 F_{(1,3,10)_{\overline\Sigma}} =\bar \sigma \left[ m_\Sigma + 
\eta_A (3 a + 2 b) \right]
&=& 0\;,
\label{efeqs}
\end{eqnarray}

\noindent
where 
\begin{eqnarray}
s = \langle (1,1,1)_S \rangle  \quad , \quad
a = \langle (1,1,15)_A \rangle \quad , \quad
b = \langle (1,3,1)_A \rangle \quad ,\nonumber \\
\sigma = \langle (1,3,10)_\Sigma \rangle \quad , \quad 
\bar\sigma = \langle (1,3,\overline{ 10})_{\overline\Sigma}
\rangle \quad .
\label{vevnames}
\end{eqnarray}

\noindent 
All other fields have zero vacuum expectation values (see
Appendix A). Notice that the choice of the two chains of 
symmetry breaking depends on the ratio of $a$ and $b$. In 
both cases it is assured that  $s$ is the largest VEV.

Next, we can imagine two possibilities:
\begin{description}
\item[(a)] $s\gg a \gg \sigma= \bar\sigma \gg b \simeq \sigma^2/s$ ,
\item[(b)] $s \gg b \gg \sigma = \bar \sigma \gg a \simeq \sigma^2/s$ ,

\end{description}

\noindent 
which correspond precisely to chains (a) and (b) 
respectively (notice that $\sigma = \bar\sigma$ is necessary for
D-flatness, see Appendix A). This can be achieved by paying the usual
price of fine-tuning,  an issue which is beyond the scope of this paper.
In case (a) this implies

\begin{equation}
m_A + 2 \lambda s \simeq {\sigma^2 \over a} \ll s\;,
\label{finea}
\end{equation}

\noindent which then ensures
\begin{equation}
b \simeq {\sigma^2 \over s} \ll \sigma
\label{finea2}
\end{equation}

\noindent 
(it is important to keep in mind that $b$ can never vanish). In 
case (b), of course, the conditions are  obtained interchanging 
the roles of $a$ and $b$, i.e. $m_A-3\lambda s\simeq 
\sigma^2/b\ll s$.

A comment is in order. In the F flatness conditions (\ref{efeqs})
we ignore the fields in {\bf 16} and {\bf 10}
 dimensional representations. This  
is justifiable for the Standard Model non-singlet fields, but not for 
$\tilde{\nu^c}$ in {\bf 16}. 
However, as we already noticed, $\tilde{\nu^c}$ 
is coupled to the SU(5) singlet in $\overline\Sigma$ which
gets a VEV $\bar\sigma$. The F-equations will then give

\begin{equation}
F_{\tilde{\nu^c}}= \sigma \langle \tilde{\nu^c}\rangle =0
\end{equation}

\noindent
guaranteeing a vanishing VEV for $\tilde{\nu^c}$. This is a 
general feature of theories with a renormalizable see-saw 
mechanism, as shown in \cite{amrs99}. We readdress this 
important issue below, when we discuss the fate of R-parity.

A few words about the low-energy sector of the theory. Besides the 
above fields we also need the usual 16-dimensional representations, 
which give the standard model fermions and sfermions and 
representations containing the electroweak Higgs doublets. 
For the sake of minimality we choose for the latter 
the 10-dimensional representation. We leave their number 
open in order to be as general as possible. It is well known that 
the minimal theory with just one ${\bf 10}$ predicts wrong fermion 
masses and no CKM mixing if one restricts oneself to the tree level 
and/or renormalizable interactions. The higher dimensional 
operators can easily introduce small flavor mixings and correct the 
light quark mass ratios. Recently it has been 
emphasized that the radiative corrections in supersymmetric models 
can also do the job \cite{bdm98}. We present our results for both 
the minimal model and the model with two $10$-dimensional Higgs 
supermultiplets, since the latter certainly works.

\vspace{0.3cm}

\section{Particle Spectrum.}   

Since we wish to determine the scales of symmetry breaking, we 
need the precise particle spectrum of both light ($\sim M_W$) 
and heavy ($\gg M_W$) states. The point here is that in 
supersymmetric theories one can not rely on the survival (or 
extended) principle, since the lack of cubic terms in the 
superpotential for many representations naturally suppresses 
many particle masses \cite{abmrs99}. More precisely, if there is an 
intermediate scale $M_I$ besides the unification scale $M_X$, 
then one finds the effective quartic couplings suppressed by 
$M_I/M_X$ and there are a number of states whose masses become 
of the order of $M_I^2/M_X$ instead of $M_I$. This of course 
has a strong impact on the unification predictions. This will 
be made manifest in the examples discussed below.

\vspace{0.3cm}
{\bf Case (a)} \hspace{0.5cm}

It is a simple exercise to show the well-known fact that all 
the states in {\bf 54} become super heavy. So do most of the states 
in {\bf 45}, {\bf 126} and $\overline{\bf 126}$. The only states which do not 
pick up a mass of order $ s \simeq M_X$ in {\bf 126} and 
$\overline{\bf 126}$ are the SU(2) triplets, namely 
$(3,1,\overline{10})_\Sigma$, $(1,3,10)_\Sigma$, 
$(3,1,10)_{\overline\Sigma}$, and $(1,3,\overline{10})_{\overline\Sigma}$. 
As mentioned above we must fine-tune the mass of the field 
$(1,1,15)_A$ ($m_A + 2 \lambda s \simeq 0$).

Next, we switch-on $a \simeq M_C$ and $\sigma \simeq M_R$ with 
$a\gg \sigma$. As is clear from (\ref{chain}), $a \simeq M_C$ 
breaks PS down to LR; it is the color singlet component in 
$(1,1,15)_A$ that gets a VEV. Due to the supersymmetric version of 
the Higgs mechanism, color triplets in $(1,1,15)_A$ get a mass of 
order $a$, but the color octet remains in principle much lighter. 
Due to the absence of the trilinear term $A^3$, it can only 
get a mass of order $M_C^2/M_X$ (through the mixing with 
the color octet in $(1,1,20)_S$) or $M_R^2/M_C$ (through 
the fine-tuning condition $m_A+2\lambda s\simeq\sigma^2/a$).

Regarding the fields in $\Sigma $ and $\overline{\Sigma}$, notice 
first that the neutral component  in $(1,1,15)_A$ is a parity 
(charge conjugation) odd super-Higgs. In other words, the 
left-handed triplets $(3,1,\overline{10})_\Sigma $ and
$(3,1,10)_{\overline\Sigma}$ get masses 
$m_\Sigma - 3 a \eta_A \simeq M_C$ and decouple for lower 
energies. Similarly, the color triplet and sextet fields 
in their right-handed counterparts $(1,3,10)_\Sigma$ and 
$(1,3,\overline{10})_{\overline\Sigma}$ get masses $\sim M_C$. 

Except for the doubly-charges states $\delta_c^{++}$ 
and $\bar\delta_c^{++}$, the rest of the fields in the 
color singlet components for the above representations 
get masses $\sim M_R$ through the super-Higgs  mechanism. The 
$\delta_c^{++} $ and $\bar\delta_c^{++}$ fields, much as 
in LR supersymmetric models \cite{amrs98}, pick up their 
mass only through $b\simeq \sigma^2/s$.

The masses are summarized in Table 1 below.

\begin{center}
 \framebox{\begin{tabular}{l|l}
\hspace{2cm}State & \hspace{1cm} Mass \\
\hline 
 & \\
 \begin{tabular}{l}
all of $S$ in {\bf 54} \\
all of $A$ in {\bf 45}, except $(1,1,15)_A$\\
all of $\Sigma$ in {\bf 126} + $\overline\Sigma$ in $\overline{\bf 126}$,
 except  $SU(4)_C$ decuplets
\end{tabular}
&  $\sim M_X$ \\
 & \\
\hline
 & \\
 \begin{tabular}{l}
$(3,1,\overline{10})_\Sigma$  + $(3,1,{10})_{\overline\Sigma}$  \\
color triplets and sextets of $(1,3,10)_\Sigma $  and
 $(1,3,\overline{10})_{\overline\Sigma}$  \\
color triplets of $(1,1,15)_A$
\end{tabular}
&  $\sim M_C$ \\
 & \\ \hline & \\
 \begin{tabular}{l}
$(\delta_c^0-\overline\delta_c^0), \delta_c^+,\overline\delta_c^+  $\\
\hspace{0.5cm} from the color singlets of $(1,3,10)_\Sigma$
  and $(1,3,\overline{10})_{\overline\Sigma}$ 
\end{tabular}
&  $\sim M_R$ \\
& \\ \hline & \\
 color octet and singlet of $(1,1,15)_A$
&  $\sim M_1 \equiv Max\left[ {M_R^2\over M_C}, {M_C^2\over M_X} \right]$ \\
& \\ \hline & \\
 \begin{tabular}{l}
$(\delta_c^0+\bar\delta_c^0),\delta_c^{++},\bar \delta_c^{++}  $\\
\hspace{0.5cm} from the color singlets of $(1,3,10)_\Sigma$ 
 and $(1,3,\overline{10})_{\overline\Sigma}$ 
\end{tabular}
&  $\sim M_2 \equiv {M_R^2 / M_X}$ \\
 & \\
\end{tabular}}

\end{center}

\vspace{0.5cm}

\noindent
{\bf Table 1}: Mass spectrum for the symmetry breaking chain
SO(10)$\stackrel{M_X}{\rightarrow}$SU(2)$_L\times $SU(2)$_R \times$ SU(4)$_C$$
\stackrel{M_C}{\rightarrow}$ 
SU(2)$_L\times$ SU(2)$_R \times$ U(1)$_{B-L}
\times$ SU(3)$_C$ $\stackrel{M_R}{\rightarrow}$ SU(2)$_L \times$ U(1)$_Y
\times$ SU(3)$_C$.

\vspace{0.5cm}

Notice the `light' states in the two last rows of the type 
we discussed above as the violation of the survival principle. 
Their small masses are the product of the lack of renormalizable 
interactions and fine-tuning conditions.

On top of this there are of course three generations of 
fermions and sfermions which lie at $M_W$ or below. The right-handed 
neutrino supermultiplet is at $M_R$, but it does not 
affect the running. In the minimal model the 10-dimensional 
Higgs multiplet splits into a light ($\sim M_W$ or supersymmetry 
breaking scale $M_S$) bidoublet, 
which provides the two MSSM doublets, and the 
superheavy ($\sim M_X$) color triplet and antitriplet. 
If there are two {\bf 10}s, then besides the two light doublets 
and all the superheavy colored states we will have 
two more doublet superfields with masses $>>M_W$. The 
range for the heavy doublet masses depend on one's scenario 
for generation of the weak mixing angles. If these are to 
arise at the tree level from bi-doublet mixing induced by the 
vev of the (1,3,1) submultiplet of the ${\bf 45}$, then one 
needs more than one 10-plet and it is necessary to assume 
$M_H\sim b$ (where $M_H$ is the coefficient of the ${\bf 10}^2$ 
term in the superpotential), while the coupling $\lambda_H\sim 
b/M_X$ (where $\lambda_H$ is the coefficient of the 
${\bf 54\; 10}^2$ term in the superpotential), in which case 
the extra doublets have mass $\sim M_R^2/M_X$. Else the 
mixing is negligible and the doublets are superheavy so that 
the theory effectively reduces to the case with a single 
${\bf 10}$. One may obtain realistic mixing even with a single
10-plet if radiative corrections due to soft terms are 
appropriate \cite{bdm98}. 

\vspace{0.3cm}

{\bf Case (b)} \hspace{0.5cm} Let us now turn to case (b) in 
(\ref{chain}). This implies interchanging the values of the 
{\bf 45} VEVs $a$ and $b$. The analysis proceeds along the same
lines, so we just present the particle spectrum in Table 2.

\begin{center}
\framebox{ \begin{tabular}{l|l}
\hspace{2cm}State & \hspace{1cm} Mass \\
  \hline & \\ 
 \begin{tabular}{l}
all of $S$ in {\bf 54} \\
all of $A$ in {\bf 45}, except $(3,1,1)_A + (1,3,1)_A$\\
all of $\Sigma$ in {\bf 126} + $\overline\Sigma$ in $\overline{\bf 126}$,
 except  $SU(4)_C$ decuplets
\end{tabular}
&  $\sim M_X$ \\
 & \\ \hline & \\
 \begin{tabular}{l}
$(3,1,\overline{10})_\Sigma$  + $(3,1,{10})_{\overline\Sigma}$  \\
$(1, ^{\;0}_{\;-},10)_\Sigma$  and
 $(1,^{\,+}_{\;0},\overline{10})_{\overline\Sigma}$  \\
$\omega_c^\pm$ from $(1,3,1)_A$
\end{tabular}
&  $\sim M_{R}$ \\
 & \\ \hline & \\
 \begin{tabular}{l}
color triplets and singlets from  \\
\hspace{0.5cm}$(1,+,10)_\Sigma$  and
$(1,-,\overline{10})_{\overline\Sigma}$ 
\end{tabular}
&  $\sim M_C$ \\
 & \\ \hline & \\
 $(3,1,1)_A$ &
 $\sim M_1 \equiv  Max\left[{ M_R^2\over M_X}, {M_C^2\over M_R} \right]$\\
 & \\ \hline & \\
 \begin{tabular}{l}
color sextets from \\
 \hspace{0.5cm}$(1,+,10)_\Sigma$  and
$(1,-,\overline{10})_{\overline\Sigma}$ 
\end{tabular}
&  $\sim M_2 \equiv  {M_C^2 / M_X}$\\
&  
\end{tabular}}

\end{center}

\vspace{0.5cm}

\noindent
{\bf Table 2}: Mass spectrum for the symmetry breaking chain
SO(10)$ \stackrel{M_X}{\rightarrow}$SU(2)$_L\times $SU(2)$_R \times$ SU(4)$_C$$ \stackrel{M_{R}}
{\rightarrow}$
SU(2)$_L\times$ U(1)$_R\times$ SU(4)$_C  \stackrel{M_C}{\rightarrow}
$ SU(2)$_L \times$ U(1)$_Y
\times$ SU(3)$_C$ .
The states in $(1,3,10)$ and $(1,3,\overline{10})$ were decomposed 
according to their $T_{3R}$ number, for example $(1,+,10)$ denotes 
the component of $(1,3,10)$ with $T_{3R} = +1$, etc. 

\vspace{0.5cm}

Notice again the survival of the fittest principle. 
These are the color sextets and left-handed triplet states.
Again their small masses are due to the lack of renormalizable 
interactions or the fine-tuning condition 
($ m_A - 3 \lambda s \simeq \sigma^2/b \simeq M_C^2/M_R$).

In this case the heavy bi-doublets will have masses of order 
$b\sim M_R$ provided $M_H$, $\lambda_H$ are again chosen small. 
Else one may again use a single 10-plet and radiative effects.

\vspace{0.3cm}

\section{Unification of gauge couplings}
 
Armed with the complete spectrum of the physical states we are now 
ready to perform the analysis of the unification constraints. 
Notice that this is a precise analysis without any {\it ad hoc}
assumptions  made regarding  the spectrum at intermediate scales. 
At this stage we wish to have a rough qualitative estimate 
of the new mass scales and thus it is only appropriate to perform 
this at the one-loop level. In this way we check whether the 
assumptions (a) and (b) are consistent with the low-energy 
values of the coupling constants.

We define the running coefficients as usual,

\begin{equation}
{1\over \alpha_i(E_1)} = {1 \over\alpha_i(E_2)} - {b_i \over 2\pi} 
\ln{E_2\over E_1}\;,
\label{running}
\end{equation}

\noindent 
where $i= 1,2,3$ stands for the normalized hypercharge,
SU(2)$_L$ and SU(3)$_C$ gauge couplings, respectively.

For the multiscale case, which we are interested in, 
eqs. (\ref{running}) become for the chain {\bf (a)}

\begin{eqnarray}
\label{gen}
{2\pi\over\alpha_i(M_Z)}={2\pi\over\alpha_U}-&[&
b_i^{(1)}ln{M_S\over M_Z}+
b_i^{(2)}ln{M_2\over M_S}+
b_i^{(3)}ln{M_1\over M_2}+\nonumber\\
&&b_i^{(4)}ln{M_R\over M_1}+
b_i^{(5)}ln{M_C\over M_R}+
b_i^{(6)}ln{M_X\over M_C}]\;,
\label{fourteen}
\end{eqnarray}

\noindent
where $M_1\equiv Max[M_R^2/M_C,M_C^2/M_X]$, $M_2 \equiv M_R^2/M_X$; 
for case {\bf (b)} one has to interchange $M_R$ and $M_C$ in
(\ref{fourteen}) as well as in the definitions of $M_1$ and $M_2$. 

\vspace{0.3cm}

Let us study the cases {\bf (a)} and {\bf (b)} defined in
(\ref{chain}) separately.

\noindent {\bf (a)} SO(10)$\stackrel{M_X}{\rightarrow}$SU(2)$_L\times $SU(2)$_R \times$ SU(4)$_C$$ 
\stackrel{M_C}{\rightarrow}$SU(2)$_L\times$ SU(2)$_R \times$ U(1)$_{B-L}
\times$ SU(3)$_C$$\stackrel{M_R}{\rightarrow}$ 
SU(2)$_L \times$ U(1)$_Y
\times$ SU(3)$_C$. The coefficients are

\begin{equation}
\begin{tabular}{c|c|c|c}
Energy range & $b_1^{(k)}$ & $b_2^{(k)}$ & $b_3^{(k)}$ \\
\hline\\
$M_Z < E < M_S$ & $-(41+n)/10$  & $ (19-n)/6$ & $7$ \\
$M_S < E < M_2$ &$ -33/5 $ &  $ -1$ & $ 3$  \\
$M_2< E < M_1$ &$ -(57 + 3\epsilon)/5$ & $-(1 + \epsilon)$ &
$3$\\
$M_1< E <M_R$  & $-(57 + 3\epsilon)/5$ & $-(1 + \epsilon)$ &
$0$ \\
$M_R < E < M_C $&$ -(45 + 3\epsilon)/5$ & $-(1 + \epsilon)$ &$0$ \\
$M_C < E < M_X $& $-(191 + 3\epsilon)/5$ & $-(41 + \epsilon)$ & $-34$\\
\end{tabular}
\label{coefa}
\end{equation}

\noindent
where $\epsilon$ counts the number of heavy Higgs doublets 
that kicks in at $M_R^2/M_X$ and $(n+1)$ is the 
total number of Higgs doublets below the 
supersymmetry breaking scale $M_S$. The detailed equations that follow 
from (\ref{fourteen}) specific to these models, as functions of
$\epsilon$ and $n$, are given in Appendix B.
They  can be solved to obtain the three mass scales
$M_X$, $M_C$ and $M_R$ in terms of the unification coupling
$\alpha_U$. Consistency will determine $M_1$, which is the maximum of
$M_R^2/M_C$ and $M_C^2/M_X$. Notice that we have assumed in 
(\ref{coefa}) that $M_1<M_R$, which is not necessarily true. 
However equation (\ref{coefa}) shows that only $b_1$ is 
sensitive to the scale $M_R$ (all the other coefficients remain 
constant), while only $b_3$ can feel the other scale $M_1$. 
For this reason it is irrelevant for the form of the 
renormalization group equations which of the two scales is smaller.

We have performed the calculation for
values of the coupling constants $\alpha_1(M_Z)=0.01695$,
$\alpha_2(M_Z)=0.03382$ and $\alpha_3(M_Z) = 0.119 \pm 0.003$.  

For the case with just one {\bf 10} ($\epsilon = 0$), the unification 
scale $M_X$ turns out to be independent of $\alpha_U$. $M_1$ is 
$M_R^2/M_C$ in this case, and it is found, for $n=0$

\begin{eqnarray}
\log\left({M_X \over GeV} \right) &=& 16.20 \pm 0.07 - 0.14 \log\left({M_S
\over TeV} \right) \nonumber \\
\log\left({M_C \over GeV} \right) &=& 14.43 \pm 0.07 - 0.25 \log\left({M_S
\over TeV} \right)  + 0.07 \alpha_U^{-1}\nonumber \\
\log\left({M_R \over GeV} \right) &=& 13.71 \pm 0.12 - 0.18 \log\left({M_S
\over TeV} \right) + 0.10 \alpha_U^{-1} 
\end{eqnarray}
The lowest values are obtained for larger $\alpha_3$. With $M_S
\sim 1$ TeV and $\alpha_U\sim 1/5$, we get

\begin{equation}
\log\left({M_C \over GeV} \right) \simeq 14.77 \pm 0.07 \quad
\log\left({M_R \over GeV} \right) \simeq 14.21 \pm 0.12
\end{equation}

Adding an extra 10-dimensional Higgs ($\epsilon = 1$) spoils 
the independence of $M_X$. In this case $M_1 = M_C^2/M_X$.
We get, for $n=0$

\begin{eqnarray}
\log\left({M_X \over GeV} \right) &=& 14.38 \pm 0.25 + 0.14 \log\left({M_S
\over TeV} \right) + 0.09 \alpha_U^{-1} \nonumber \\
\log\left({M_C \over GeV} \right) &=& 12.77 \pm 0.25 + 0.03 \log\left({M_S
\over TeV} \right)  + 0.15 \alpha_U^{-1}\nonumber \\
\log\left({M_R \over GeV} \right) &=& 10.25 \pm 0.46 + 0.34 \log\left({M_S
\over TeV} \right) + 0.28 \alpha_U^{-1}
\label{17} 
\end{eqnarray}

We have the interesting result of $M_X$ being {\em lower} for lower
intermediate scales $M_R$. The smallest possible value of
$(\alpha_U)^{-1}$  is dictated by proton decay 
constraints, requiring $\log(M_X/GeV) \geq 15.5$. Again,
raising the supersymmetry breaking scale has the effect of lowering
the intermediate scales, and for $M_S$ of order $1$ TeV 

\begin{equation}
\log\left({M_C \over GeV} \right) \simeq 14.63 \pm  0.17 \quad
\log\left({M_R \over GeV} \right) \simeq 13.62 \pm 0.29 
\end{equation}
The lowest possible value corresponds to $\alpha_3=0.122$, and
gives $\alpha_U^{-1}= 9.5$.

This low value of $M_R$ is very interesting for neutrino physics, 
and so we concentrate on this case, $\epsilon = 1$. Figure 1  
shows that the three scales $M_X, M_C, M_R$ increase as a 
function of $1/\alpha_U$, for $M_S = 1$ TeV. It is 
interesting to see that the requirement of proton stability 
($M_X>10^{15.5}$ GeV) guarantees the perturbativity of the 
unified coupling, $1/\alpha_U >12$.

In Figure 2 we plot our findings for $M_R$ and $M_C$ as 
functions of $M_S$ for a fixed value of $M_X = 10^{15.5}$ GeV. 
Notice that $M_R$ is almost insensitive to $M_S$ (and 
completely so for $n=1$), unlike $M_C$. Of course, the value of 
$M_S \simeq 1$ TeV is physically most interesting. Notice that 
there is some sensitivity to the precise value of $\alpha_3$.

\vspace{0.3cm}
\noindent {\bf (b)}
SO(10)$ \stackrel{M_X}{\rightarrow}$SU(2)$_L\times $SU(2)$_R \times$ SU(4)$_C$ $\stackrel{M_{R}}
{\rightarrow}$
SU(2)$_L\times$ U(1)$_R\times $SU(4)$_C  \stackrel{M_C}{\rightarrow}
$ SU(2)$_L \times$ U(1)$_Y
\times$ SU(3)$_C$.

The $b$ coefficients are now (here $\epsilon$ is the number 
of bidoublets from $10$ with mass $M_R$) 

\begin{equation}
\begin{tabular}{c|c|c|c}
Energy range & $b_1^{(k)}$ & $b_2^{(k)}$ & $b_3^{(k)}$ \\
\hline\\
$M_Z < E < M_S$ & $-(41+n)/10$  & $ (19-n)/6$ & $7$ \\
$M_S < E < M_2$ &$ -33/5 $ &  $ -1$ & $ 3$  \\
$M_2< E < M_1$ &$ -97/5$ & $-1$ &
$-2$\\
$M_1< E <M_C$  & $-97/5$ & $-3$ &
$-2$ \\
$M_C < E < M_R $&$ -81/5$ & $-3$ &$0$ \\
$M_R < E < M_X $& $-(189 + 3\epsilon)/5$ & $-(43 + \epsilon)$ & $-30$\\
\end{tabular}
\label{coefb}
\end{equation}

It can be seen that the extra bidoublets do not play a significant
role in this case, 
contributing only at scales above $M_R$. In the minimal case of
$\epsilon=0, n=0$, we get $M_1 = M_C^2/M_R$, and 

\begin{eqnarray}
\log\left({M_X \over GeV} \right) &=& 14.88 \pm 0.11 - 0.13 \log\left({M_S
\over TeV} \right) + 0.06 \alpha_U^{-1} \nonumber \\
\log\left({M_R \over GeV} \right) &=& 13.19 \pm 0.11 -0.24 \log\left({M_S
\over TeV} \right)  + 0.12 \alpha_U^{-1}\nonumber \\
\log\left({M_C \over GeV} \right) &=& 13.04 \pm 0.14 - 0.17 \log\left({M_S
\over TeV} \right) + 0.13 \alpha_U^{-1} 
\end{eqnarray}

Again, lower intermediate scales are obtained for lower $M_X$. With
$M_X \sim 10^{15.5}$ GeV,in this case, a smaller value of $M_S$  is
 preferred.  With $M_S \sim M_Z$,

\begin{equation}
\log\left({M_R \over GeV} \right) \simeq 14.47 \pm  0.12 \quad
\log\left({M_C \over GeV} \right) \simeq 14.35 \pm 0.12 
\end{equation}
the lowest value obtained for smaller $\alpha_s$,and $\alpha_U$ is now 8.7.
The case with the extra bidoublet gives essentially the same
results. 

What we observed here is that there 
can be a separation of all the scales involved, although
sometimes not very big. The presence of extra light states
(in addition to those of MSSM) below $M_X$, especially
those that have non-trivial hypercharge, tended to spoil
the succesful unification for too big separation of
scales. However, as shown in (\ref{17}), it is perfectly
possible to push scales to the low intermediate range.
Finally we remind that the results presented here are
only a rough guide, since we have not included
the two-loop and threshold effects.

\section{Phenomenological and cosmological consequences} 

We now turn to the discussion of the resulting physics 
of the theory. The most important features are the exact 
R-parity at low energies, i.e. the MSSM as an effective 
low-energy theory and the ``quasi-canonical'' see-saw form with 
a rather light neutrino.

\vspace{0.3cm}
{\bf i)} R-parity. \hspace{0.5cm} Here the situation parallels 
the one already found by some of us in the context of LR 
symmetric theories \cite{amrs98}. We showed that at the 
large scale $\langle 16 \rangle = \langle \tilde\nu_c\rangle =0$ 
and so R-parity is conserved. Next, as in any theory based 
on the renormalizable see-saw mechanism \cite{amrs99}, one can 
show that $\langle \tilde\nu\rangle = 0$ too. This is valid 
at all energies all the way to the electroweak scale and below 
(of course one has to assume $\langle\tilde\nu\rangle=0$ at 
high scale, as with other fields that break the SM symmetry).

It is worth repeating this simple, but important argument. 
It is well-known that in the MSSM we cannot break  R-parity 
spontaneously. The problem is this would imply the existence 
of the Majoron $J$, since this also breaks lepton number 
spontaneously \cite{am82}, and the $Z$ boson would then 
have an extra decay into a Majoron and its real partner. 

Now, after we integrate the heavy fields  out, we are left 
effectively with the MSSM: at low energies all the effects 
of the new scale $M_R$ must go as $1/M_R$. The same is  
true of the would-have-been Majoron $J$: its mass must 
be suppressed by $1/M_R$ and $Z$ can still decay into 
$J + partner$. Again, as in the MSSM, this is ruled out 
and we learn that R-parity can never be broken. This is 
an extremely important prediction, since then the LSP is 
absolutely stable.

\vspace{0.3cm}
{\bf ii)} Neutrino Mass. \hspace{0.5cm} One of the main motivations 
behind the SO(10) unification is the natural implementation of the 
see-saw mechanism. We chose ${\bf 126}$ and $\overline{\bf 126}$ 
Higgs multiplets in order to have a renormalizable see-saw which 
then gave us R-parity to be exact all the way down to the MSSM. 

The first thing we notice is that for both patterns of symmetry 
breaking the allowed value for $M_R$ is pretty high (usually about
$10^{14}$ GeV or higher, although it can go down to $10^{12}$ GeV, 
as in (\ref{17})). Second, supersymmetry helps us not to have
uncontrollable VEVs from the left-handed triplet VEVs, and 
indeed the supersymmetry breaking effects are negligible for 
such high $M_R$\cite{amrs99}.

Further, as we emphasized in \cite{amrs99}, the exact form of the
see-saw is rather model dependent in supersymmetric SO(10) theories. 
The canonical form for the see-saw for our model is spoiled by 
effective non-renormalizable terms in the superpotential of the form

\begin{equation}
{1 \over M_X} \bar\Delta \Phi^2 \bar\Delta_c
\label{noncan}
\end{equation}

\noindent
(where $\bar\Delta$ and $\bar\Delta_c$ are left and right-handed 
triplets respectively, from $\overline{\bf 126}$ and $\Phi$  is a
bidoublet 
from {\bf 10}) which arise once the heavy ${\bf 54}$ fields get 
integrated out. Such terms then produce nonvanishing VEVs for the
left-handed triplets since they get generated from the term above with
$m_\Sigma \Delta \overline{\Delta}$
to give

\begin{equation}
\langle \bar\Delta \rangle = {{\langle \bar\Delta_c \rangle} \over 
{M_X m_\Sigma}} 
\langle \Phi \rangle^2
\equiv
\epsilon {{\langle \Phi \rangle^2} \over 
{\langle \bar\Delta_c \rangle}}
\end{equation}

\noindent
where $\epsilon \equiv {{\langle \bar\Delta_c \rangle^2}/ 
{M_X m_\Sigma}}$ can be anywhere between 1 and $10^{-4}$ 
in the above chains of symmetry breaking.

Thus the formula for neutrino masses is 

\begin{equation}
m_\nu \approx ( f^2 \epsilon - h^2_D ) 
{{\langle \Phi \rangle^2} \over m_{\nu_R}}
\label{seesa}
\end{equation}
where $f$ is the coupling of triplets and right-handed neutrinos.

We see now that the $\epsilon$ term can not be neglected 
compared to the canonical term coming from the neutrino 
Dirac masses, at least for the lighter generations. This 
might be a welcome addition to the otherwise
rather constrained scenarios\cite{cein99}.

Of course, in some specific models one can have this new 
non-canonical term in the see-saw formula vanishing. For 
example, this is what happens if the Dimopoulos-Wilczek 
mechanism \cite{dw83} is employed to solve the doublet-triplet 
splitting problem. Let us briefly repeat the argument 
\cite{amrs99}. As mentioned above, the form (\ref{noncan}) 
is obtained from the terms ($H$ is the $10$-dimensional Higgs) 

\begin{equation}
\lambda_S S \Sigma^2 + \lambda_H S H^2 +{m_S\over 2}S^2
\end{equation}

\noindent
after integrating out the heavy $S$, for which the presence 
of the term $\lambda_H S H^2$ is crucial. However, to solve 
the doublet-triplet problem this same term must be absent, 
which is obtained for example with an additional $Z_3$ 
symmetry \cite{babu94}. The Dimopoulos-Wilczek mechanism 
is then implemented by the new terms (among others) 

\begin{equation}
\lambda_{H'} S H'^2 + \lambda_{HH'} H A' H' \;,
\end{equation}

\noindent
where $H$ and $H'$ are ${\bf 10}$'s and $A'$ a new ${\bf 45}$ with 
the vev $<A'>=diag(a',a',a',0,0)\times\tau_2$. 

Since the $\lambda_H S H^2$ term is forbidden, only the 
SU(2)$_L$ doublets from $H$ (but not from $H'$) have zero mass. 
So the Dimopoulos-Wilczek mechanism at the same time solves 
the doublet-triplet splitting problem and gives the 
canonical see-saw. The price to pay is however the ad-hoc addition 
of a discrete symmetry and the inclusion of more extra fields.

\vspace{0.3cm}
{\bf iii)} Proton decay and related issues. \hspace{0.5cm} 
The usual lore is that the unification constraints favor 
the minimal supersymmetric SU(5) grand unified theory. One 
gets the single scale GUT breaking $M_X \simeq 2\times 10^{16}$
GeV, which makes the usual $d=6$ contribution to proton 
decay quite small:

\begin{equation}
\tau (p \to \pi^0 e^+) \geq 10^{36 \pm 1.5} yr.
\end{equation}

\noindent
and the dominant decay mode $p \to K^+ \bar\nu$ comes from the $d=5$
operators\cite{hmy91}.
It is also known that in SO(10) models, due to misalignment
of the Higgs triplet and doublet Yukawa couplings and depending 
on the details of the underlying flavor theory, other
decay modes to kaons may be dominant, such as 
$p \to K^0 l^+$ \cite{babu}. 

However, what we see in the types of models studied here, is
that unification through alternate symmetry breaking channels is quite
possible. Moreover even the ``smoking gun'' signature of supersymmetric
models with kaon production in proton decay, may not be dominant. Namely,
we saw that, contrary to the common belief, lowering of the intermediate
scale $M_R$ can cause {\it lowering} of the grand unified scale
(at least in the case with more than one 10 Higgs multiplet).
This then opens the possibility that the gauge boson mediated proton
decay may be the dominant one and thus we could even have the possibility
where in a supersymmetric model the dominant decay products be 
pions instead of kaons \footnote{An important, although not dominant 
contribution of the $p\to\pi^0 e^+$ mode has been found in
\cite{wilczek}, but there it was produced by $d=5$ operators.}!

Let us also comment on other higher dimensional operators. 
We know that the dangerous $d=4$ proton decay is absent, 
since R-parity is exact even at  $E< M_W$. Of course, R-parity 
cannot guarantee the absence  of higher dimensional operators 
which break the baryon and lepton number. Since 
${\bf 16}^4$ is invariant under SO(10), one can have 

\begin{equation}
\Delta W_{effective} = C{Q Q Q L \over M} + ...
\label{qqql}
\end{equation}

Even if $M= M_{Pl}$, unless $C$ is extremely small, the above 
interaction would be dangerously large. Of course, there is no 
proof that it must be present, but still, it would be nice to 
have some explanation for its absence.

If one  dislikes ad-hoc discrete symmetries, as we do (in spite 
of the euphemism that they can come from superstrings), one 
natural possibility is to go to $E(6)$ GUT. In $E(6)$ the 
basic representation is $27 = 16_{SO(10)} +
10_{SO(10)} + 1_{SO(10)}$, and the dangerous interaction 
$(27)^4$ is not allowed. Of course, one must then split 
naturally the Higgs in $10_{SO(10)}$ from the new states 
$D$ and $D^c$ (down-quark like) in $10$. The study of this 
is beyond the scope of this work and is related to the 
construction of the $E(6)$ grand unified theory.

Of course, even without going beyond SO(10) or including 
nonrenormalizable operators, one gets proton decay mediating 
$d=5$ operators (\ref{qqql}) as usual after integrating 
out the heavy color triplets and antitriplets, i.e. 
$(1,1,6)$ from ${\bf 10}$ (the same contribution as in SU(5)), 
but also from $\overline{\bf 126}$ \cite{wilczek}. 

\vspace{0.3cm}

\section{ Summary }

There are a number of well-known reasons that make SO(10) 
a popular grand unified theory. The two main ones are the 
grouping of families of fermions in  16-dimensional spinor
representations, and the natural incorporation of the see-saw
mechanism.  The supersymmetric extension of the theory provides
yet another important rationale, namely, the otherwise {\it ad
hoc} discrete R-parity symmetry of the MSSM becomes a finite
SO(10) gauge transformation. As we showed in this paper, the
symmetry remains exact throughout all the stages of symmetry 
breaking down to the electroweak scale. Thus, the low energy
effective theory is completely determined: it is the MSSM with
absolutely stable LSP. This result is a general property of the
supersymmetric see-saw mechanism and a spontaneously broken B-L
symmetry \cite{amrs99}. 

An important phenomenological prediction of the theory regards 
the proton lifetime. Remarkably enough, the theory allows for an
intermediate see-saw scale at the expense of lowering the
unification scale compared to the single-step breaking of
supersymmetric GUTs. As argued in \cite{abmrs99}, the 
existence of intermediate scales is due to the violation of the
so-called survival principle, i.e. to the fact that there 
is a number of supermultiplets whose masses lie below the 
corresponding scales of symmetry breaking. The typical value 
of their masses is $\sim M_R^2/M_X$. We find that the lowest
possible values of $M_R$ are of order $  10^{13-14} $GeV,
and with $M_X \sim 10^{16}$ GeV,  these states turn out to 
be out of experimental reach. 

Contrary to the usual belief, the low unification scale 
allows for the interesting possibility of proton decay being
dominated by the usual $d=6$ mode characteristic of ordinary 
GUTs, $p \to e^+\, \pi^0$. Now, this is tied up, as we said, 
to an intermediate see-saw scale in the range $10^{13-14}
$GeV. This is encouraging for the neutrino mass expectations. 
However, we must admit that (unlike the often present claims 
in the literature), one cannot really predict  neutrino 
masses. This is clear from the central formula (\ref{seesa}), 
which shows that besides the usual canonical term proportional 
to the squared Dirac Yukawa coupling, there is another piece 
which cannot be predicted by the SO(10) symmetry itself. Clearly,
there is enough freedom to accomodate the solar and atmospheric
neutrino data, but no way of making honest predictions. 

\vspace{0.5cm}

{\it Acknowledgements.}

We are grateful to Gia Dvali for many discussions. 
The work of G.S. is supported  by EEC under the TMR contract 
ERBFMRX-CT960090, that of A.M. by CDCHT-ULA Project No. 
C-898-98-05-B, that of A.R. by the U.S. Department of Energy grant
DE-FG02-97ER-41036 and that of B.B. by the Ministry of Science 
and Technology of the Republic of Slovenia and by the Packard 
Foundation 99-1462 fellowship. 

\section*{Appendix A}

We will denote SO(10) indices by $i,j,..$, SO(6) indices by 
$a,b,...$, and SO(4) indices by $\alpha, \beta,..$. With this 
convention, the field $S$ in the  {\bf 54} representation is 
represented by the symmetric, traceless second order tensor 
$S_{ij}$; the field in the {\bf 45} representation, $A$, by the 
antisymmetric second order $A_{ij}$; and the fields $\Sigma$ and
$\overline\Sigma$ in the {\bf 126} and $\overline{\bf 126}$
representations by the fifth-order antisymmetric tensors 
$\Sigma_{ijklm}$ and $\overline \Sigma_{ijklm}$, with self-
and anti-self-duality conditions 

$$
\Sigma_{ijklm} = {i \over 5!} \epsilon_{ijklmopqrs} \Sigma_{opqrs}
\quad , \quad \overline \Sigma_{ijklm} = -{i \over 5!}
\epsilon_{ijklmopqrs} 
 \overline\Sigma_{opqrs}\quad .
$$

The superpotential is

\begin{eqnarray}
W &=&  {m_S \over 2} {\rm Tr} S^2 + 
{\lambda_S \over 3} {\rm Tr S^3} + 
{m_A \over 2} {\rm Tr} A^2 
+ \lambda {\rm Tr} A^2 S  \nonumber \\
&+& m_\Sigma  \Sigma \overline\Sigma + 
\eta_S  \Sigma^2 S + 
\overline\eta_s \overline\Sigma^2 S +
\eta_A \Sigma\overline\Sigma A  \quad ,
\label{Asuperpot}
\end{eqnarray}

where

$$
\Sigma\overline\Sigma A = {1 \over 4!} \Sigma_{ijklm} 
\overline\Sigma_{ijklp} A_{mp}
\quad ,\quad \Sigma\overline\Sigma  = \Sigma_{ijklm}
\overline\Sigma_{ijklm} \quad
, \quad \Sigma^2 S = {1 \over 4!} \Sigma_{ijklm}\Sigma_{ijklp} S_{mp}
\quad .
$$

The F-terms are then

\begin{eqnarray}
(F_A)_{ji} &=& m_A A_{ij}  + \lambda (A_{il} S_{lj} + S_{il} A_{lj})
+ {\eta_A \over 4!} [\Sigma_{klmpj} \overline\Sigma_{klmpi} -
\Sigma_{klmpi} \overline\Sigma_{klmpj}] \;, \nonumber\\
& & \nonumber \\
(F_S)_{ij} &=& m_S S_{ij} + \lambda_S [ (S^2)_{ij} - {\delta_{ij}
 \over 10} {\rm Tr} S^2 ] + \lambda [ (A^2)_{ij} - {\delta_{ij}
 \over 10} {\rm Tr} A^2]
\nonumber\\ &+& {\eta_S \over 4!} \Sigma_{klmpi} \Sigma_{klmpj} +
{\overline\eta_S \over 4!} \overline\Sigma_{klmpi}
 \overline\Sigma_{klmpj}\;, 
\nonumber\\
& & \nonumber \\
{(F_{\overline\Sigma})}_{ijklm} &=& m_\Sigma \Sigma_{ijklm} + {\eta_A} 
[ A_{ip} \Sigma_{pjklm} + {i \over 4!} 
\epsilon_{ijklmopqrs} A_{ot} \Sigma_{tpqrs} ] \nonumber  \nonumber\\
&+& {2 \overline \eta_S } [ S_{ip} \overline\Sigma_{pjklm} 
+ {i \over 4!} \epsilon_{ijklmopqrs}
 S_{ot} \overline\Sigma_{tpqrs} ]\;, \nonumber \\
 & & \nonumber \\
{(F_{\Sigma})}_{ijklm} &=& m_\Sigma \overline\Sigma_{ijklm} + {\eta_A} 
[ A_{ip} \overline\Sigma_{pjklm} - {i \over 4!} 
\epsilon_{ijklmopqrs} A_{ot} \overline\Sigma_{tpqrs} ] \nonumber \\
&+& {2  \eta_S } [ S_{ip}  \Sigma_{pjklm} 
- {i \over 4!} \epsilon_{ijklmopqrs}
 S_{ot} \Sigma_{tpqrs} ]\;. 
\end{eqnarray}

The VEVs for $S$ and $A$ that ensure the required symmetry 
breaking SO(10) $\to$ SU(2)$_L\times$SU(2)$_R\times$SU(4)$_C$ are

\begin{equation}
S= s \, {\rm diag}(1,1,1,-3/2,-3/2) \times 1_2 \quad ,\quad 
A = {\rm diag}(a,a,a,b,b) \times \tau_2\quad .
\label{sanda}
\end{equation}

The VEV of $\Sigma$ and $\overline \Sigma$ are a bit more complicated. 
In the basis where  the Cartan generators are diagonal, the 
SU(2)$_L\times$U(1)$_Y \times$SU(3)$_C$ singlets in $\Sigma$ 
and $\overline \Sigma$ get VEVs:

\begin{equation}
\langle \Sigma'_{13579}\rangle = \sigma \quad , \quad
\langle \overline\Sigma'_{246810}\rangle = \bar\sigma\quad .
\label{sigandsigbar}
\end{equation}

\noindent
One then goes to  the basis in which (\ref{sanda}) is valid, 
to find the $2^5$ components of 
$\Sigma$ and $\overline \Sigma$ that get a nonvanishing VEV. 
They have the first index equal to 1 or 2; the second to 3 or 
4, etc. Each have a VEV $2^{-5/2} \sigma$, with a factor of $-i$ 
for each even index in $\Sigma$ and a factor of $+i$ for each 
even index in $\overline\Sigma$.

The VEV directions are chosen to get the required symmetry breaking, 
and of course are not the only possible ones. To make sure that 
the vacuum chosen is not connected to the undesired ones by flat 
directions, one has to examine the mass spectrum: the flat directions 
will be associated with massless excitations. It is clear from the 
analysis in Section III that no such states exist for our vacuum.

The VEV equations are then written as

\begin{eqnarray}
\sigma [ m_\sigma + \eta_A ( 3 a + 2 b)] = \bar \sigma
 [ m_\sigma + \eta_A ( 3 a + 2 b)] & = &  0 \\
 & \nonumber \\
(m_A + 2 \lambda s) a + {\eta_A \over 2} \sigma \bar \sigma & = & 0 \\
& \nonumber \\
(m_A - 3 \lambda s) b + {\eta_A \over 2} \sigma \bar\sigma & = & 0 \\
& \nonumber \\
s [m_S - {\lambda_S \over 2} s] + {2 \over 5} \lambda (a^2 - b^2) &=& 0
\end{eqnarray}

Then if one fine-tunes $(m_A + 2 \lambda s) a \sim M_R^2$,
the first pattern {\bf (a)} of symmetry breaking is obtained

\begin{eqnarray}
SO(10) &\stackrel{\langle S \rangle}{\longrightarrow}&
SU(2)_L\times SU(2)_R \times SU(4)_C 
\stackrel{\langle A \rangle}{\longrightarrow}
SU(2)_L\times SU(2)_R \times SU(3)_C \times U(1)_{BL} 
\nonumber \\ &\stackrel{\langle \Sigma \rangle}{\longrightarrow}&
SU(2)_L \times SU(3)_C \times U(1)_Y
\end{eqnarray}
with
\begin{equation}
M_X = s \sim m_S \sim m_A \; ; \quad 
M_C = a \sim m_\Sigma \; ; \quad 
M_R^2 = \sigma \bar \sigma 
\end{equation}
and  $b \sim M_R^2/M_X$. 

The second pattern {\bf(b)} is obtained by instead fine-tuning 
$(m_A - 3 \lambda s) a \sim M_R^2$. Then

\begin{eqnarray}
SO(10) &\stackrel{\langle S \rangle}{\longrightarrow}&
SU(2)_L\times SU(2)_R \times SU(4)_C \stackrel{\langle \Sigma, 
\overline\Sigma \rangle}{\longrightarrow}
SU(2)_L\times U(1)_R \times SU(4)_C 
\nonumber \\ &\stackrel{\langle A \rangle}{\longrightarrow}&
SU(2)_L \times SU(3)_C \times U(1)_Y
\end{eqnarray}

\begin{equation}
M_X = s \sim m_S \sim m_A \; \gg
M_R^2 = \sigma \bar \sigma \; \gg  \quad M_C = b \sim m_\Sigma 
\end{equation}
and  $a \sim M_C^2/M_X$. 

The contribution of the representations S(54), A(45), 
$\Sigma$(126) and ${\overline{\Sigma}}$ to the D-terms 
is as follows:

\begin{eqnarray}
D_{ij} =(-ig_U) ( 
2 S^*_{k[i} S_{j]k} - 
2 A^*_{k[i} A_{j]k} + 
5 \Sigma^*_{klmn[i}\Sigma_{j]klmn} + 
5 {\overline{\Sigma}}^*_{klmn[i}\overline{\Sigma}_{j]klmn} + 
...)\;.
\end{eqnarray}

The contributions of the two index representations S and A may 
be written as commutators 

\begin{eqnarray}
D_{ij} = (-ig_U) ( 2 [S^{\dagger}, S] + 
2 [A^{\dagger},A] + ...)_{ij}\;,
\end{eqnarray}

\noindent
but then eq. (\ref{sanda}) for the vevs of the matrices S and A
immediately give zero for their contribution to the D terms.

Thus, it remains to see what the ${\bf 126}, {\overline{\bf 126}}$
representations give. It is most convenient to work in the 
complex basis where the Cartan generators are diagonal (indices 
$A,B=1,2...10$)  which is obtained from the usual Cartesian 
basis ($i,j=1,2...10$) by the unitary transformation 
$U=1_5 \times U_2$, where 
$U_2 = {1 \over \sqrt{2}}\left(\matrix{1 & i \cr 1 & -i \cr}\right)$ 
by $V_A = U_{Ai} V_i$. 

It is easy to see that in this basis the ${\bf 126}$ and 
$\overline{\bf 126}$ part of $D_{AB}=U_{Ai}U_{Bj}D_{ij}$ is 

\begin{eqnarray} 
D_{AB} = (-5ig_U)\left(\Sigma^*_{GCDEF} g_{G[A}\Sigma_{B]CDEF} +
{\overline{\Sigma}}^*_{GCDEF}
g_{G[A}{\overline{\Sigma}}_{B]CDEF}\right)= 
D^1_{AB} - D^1_{BA}\;,
\end{eqnarray}

\noindent 
where $g_{AB}$ is the metric in the complex basis $g=1_5\times
\tau_1$  and 

\begin{eqnarray}
D^1_{AB}= (-5ig_U)\left(\Sigma^*_{GCDEF} g_{GA}\Sigma_{BCDEF} -
{\overline{\Sigma}}^*_{GCDEF}
g_{GB}{\overline{\Sigma}}_{ACDEF}\right)
\end{eqnarray}

\noindent
can have non-vanishing vev (when the vevs of 
$\Sigma,{\overline \Sigma}$ take the form of
eq. (\ref{sigandsigbar})) only if A,B form an even-odd
conjugate pair of indices. Thus, without loss of generality, 
we have for A even and B odd

\begin{eqnarray}
<D_{AB}> = -ig_U(5!) g_{AB}(|\sigma|^2 -|{\bar{\sigma}}|^2) + 
... \;.
\end{eqnarray}
 
It follows that provided the magnitudes of $\sigma,
{\bar\sigma }$ are equal, the D terms vanish at the first two
stages of symmetry breaking specified by the vevs in eqs. 
(\ref{sanda}), (\ref{sigandsigbar}).

The absence of flat directions characterized by holomorphic
invariants formed from the fields involved in the high scale
symmetry breaking ($S,A,\Sigma,{\overline\Sigma}$)  {\it{alone }}
then follows {\it{post-hoc}} from the calculated spectra of these
fields which do not include any massless modes (aside from
those eaten by massive gauge bosons).  The usual argument 
\cite{ams98,amrs98} serves to 
invoke the protection of the soft terms against
charge and color breaking flat directions involving the light
field vevs.

\section*{Appendix B}

We now give in detail the renormalization group equations 
for each symmetry-breaking pattern.

\vskip 1cm

\noindent
{\bf Case (a)} $M_2 = M_R^2/M_X$.

For $M_1$ there are two possibilities:

\vskip 0.5cm

\noindent
1) $M_1=M_R^2/M_C$

\vskip 0.5cm  

Then (\ref{gen}) becomes

\begin{eqnarray}
\label{gen1}
{2\pi\over\alpha_i(M_Z)}={2\pi\over\alpha_U}-&[&
(b_i^{(1)}-b_i^{(2)})S+
(2b_i^{(2)}-b_i^{(4)}-b_i^{(5)})R+\nonumber\\
&&(-b_i^{(3)}+b_i^{(4)}+b_i^{(5)}-b_i^{(6)})C+
(-b_i^{(2)}+b_i^{(3)}+b_i^{(6)})X
]\;,
\end{eqnarray}

\noindent
where $Y=ln(M_Y/M_Z)$ for any scale $M_Y$. Using (\ref{coefa}) 
we get

\begin{eqnarray}
{2\pi\over\alpha_1(M_Z)}&=&{2\pi\over\alpha_U}-
\left({25-n\over 10}S+{36+6\epsilon\over 5}R+
{146\over 5}C-{215+6\epsilon\over 5}X\right)\;,\\
{2\pi\over\alpha_2(M_Z)}&=&{2\pi\over\alpha_U}-
\left({25-n\over 6}S+(2\epsilon)R+40C
-(41+2\epsilon)X\right)\;,\\
{2\pi\over\alpha_3(M_Z)}&=&{2\pi\over\alpha_U}-
\left(4S+6R+31C
-34X\right)\;.
\end{eqnarray}  

\vskip 1cm

\noindent
2) $M_1=M_C^2/M_X$

\vskip 0.5cm

\noindent
In this case (\ref{gen}) becomes

\begin{eqnarray}
\label{gen2}
{2\pi\over\alpha_i(M_Z)}={2\pi\over\alpha_U}-&[&
(b_i^{(1)}-b_i^{(2)})S+
(2b_i^{(2)}-2b_i^{(3)}+b_i^{(4)}-b_i^{(5)})R+\nonumber\\
&&(2b_i^{(3)}-2b_i^{(4)}+b_i^{(5)}-b_i^{(6)})C+
(-b_i^{(2)}+b_i^{(4)}+b_i^{(6)})X
]\;,
\end{eqnarray}

\noindent
and with numbers

\begin{eqnarray}
{2\pi\over\alpha_1(M_Z)}&=&{2\pi\over\alpha_U}-
\left({25-n\over 10}S+{36+6\epsilon\over 5}R+  
{146\over 5}C-{215+6\epsilon\over 5}X\right)\;,\\
{2\pi\over\alpha_2(M_Z)}&=&{2\pi\over\alpha_U}-
\left({25-n\over 6}S+(2\epsilon)R+40C
-(41+2\epsilon)X\right)\;,\\
{2\pi\over\alpha_3(M_Z)}&=&{2\pi\over\alpha_U}-
\left(4S+40C
-37X\right)\;.
\end{eqnarray}  

\vskip 1cm

\noindent
{\bf Case (b)} $M_2 = M_C^2/M_X$.

\noindent
We get the equivalent equations as (\ref{gen}), (\ref{gen1})
and (\ref{gen2}) just interchanging $R$ and $C$. We have again
two cases:

\vskip 0.5cm

\noindent
1) $M_1=M_C^2/M_R$

\vskip 0.5cm

\begin{eqnarray}
{2\pi\over\alpha_1(M_Z)}&=&{2\pi\over\alpha_U}-
\left({25-n\over 10}S+{112\over 5}C+{108+3\epsilon\over 5}R
-{253+3\epsilon\over 5}X\right)\;,\\
{2\pi\over\alpha_2(M_Z)}&=&{2\pi\over\alpha_U}-
\left({25-n\over 6}S+4C+(38+\epsilon)R
-(43+\epsilon)X\right)\;,\\
{2\pi\over\alpha_3(M_Z)}&=&{2\pi\over\alpha_U}-
\left(4S+8C+30R-35X\right)\;.
\end{eqnarray}  

\vskip 0.5cm

\noindent
2) $M_1=M_R^2/M_X$

\vskip 0.5cm

\begin{eqnarray}
{2\pi\over\alpha_1(M_Z)}&=&{2\pi\over\alpha_U}-
\left({25-n\over 10}S+{112\over 5}C+{108+3\epsilon\over 5}R
-{253+3\epsilon\over 5}X\right)\;,\\
{2\pi\over\alpha_2(M_Z)}&=&{2\pi\over\alpha_U}-
\left({25-n\over 6}S+(44+\epsilon)R
-(45+\epsilon)X\right)\;,\\
{2\pi\over\alpha_3(M_Z)}&=&{2\pi\over\alpha_U}-
\left(4S+8C+30R
-35X\right)\;.
\end{eqnarray}

\begin{figure}
\centerline{\psfig{figure=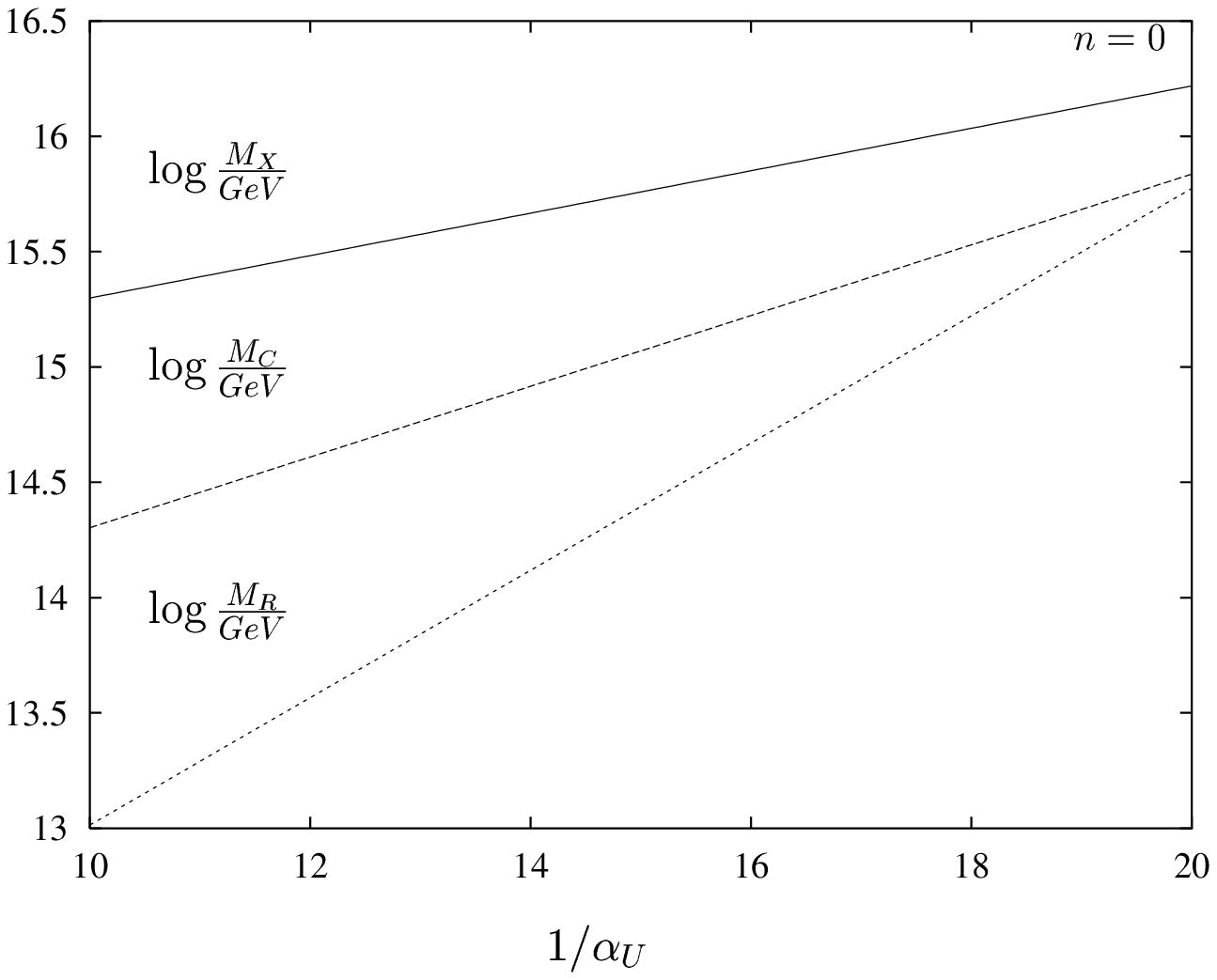,height=7cm}
\psfig{figure=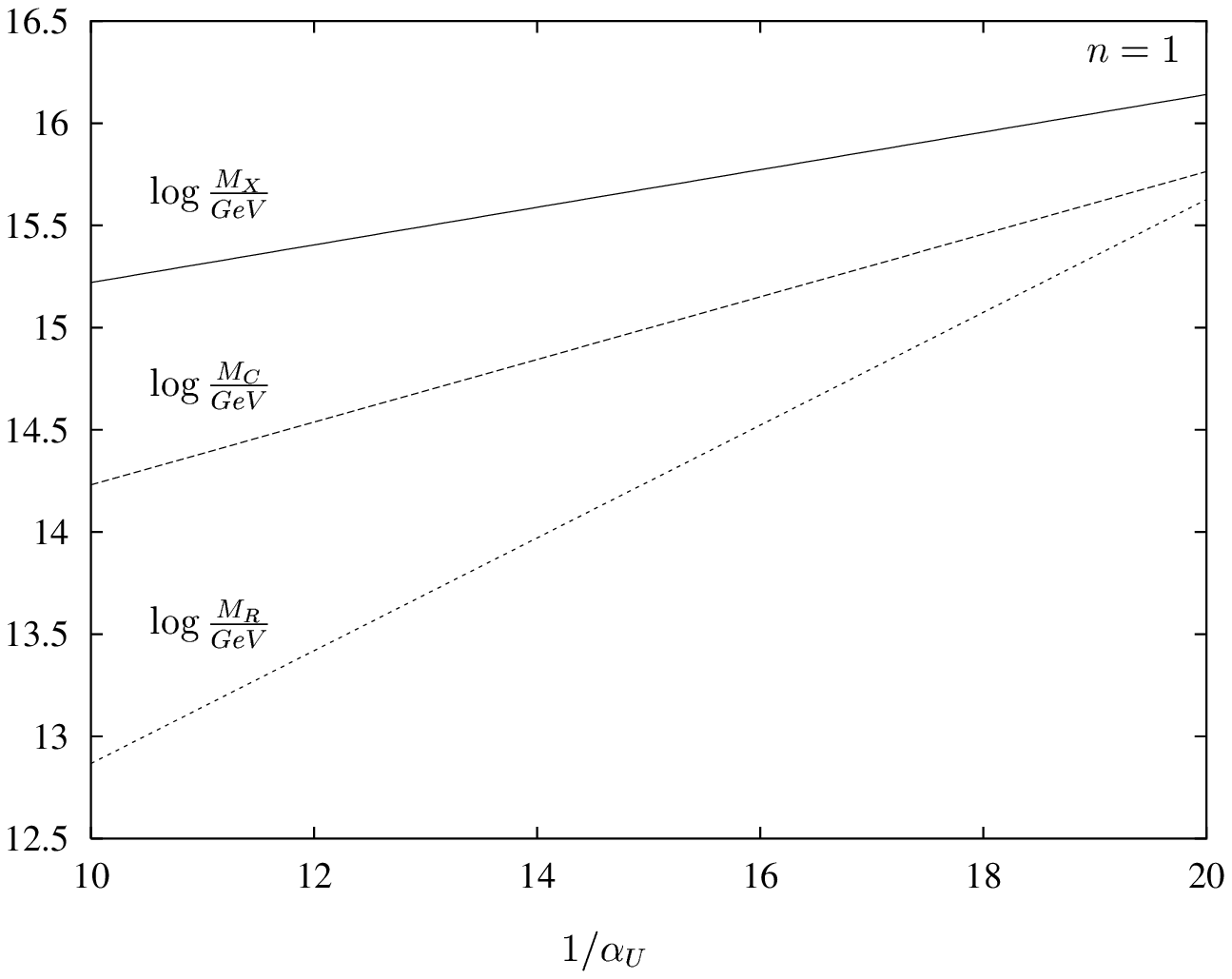,height=7cm}}
\caption{$M_C$, $M_R$ and $M_X$ vs. $1/\alpha_U$ for Chain A, 
with $M_s = 1$ TeV and $\epsilon = 1$,  for $n=0$ and $1$.}
\end{figure}

\begin{figure}
\centerline{\psfig{figure=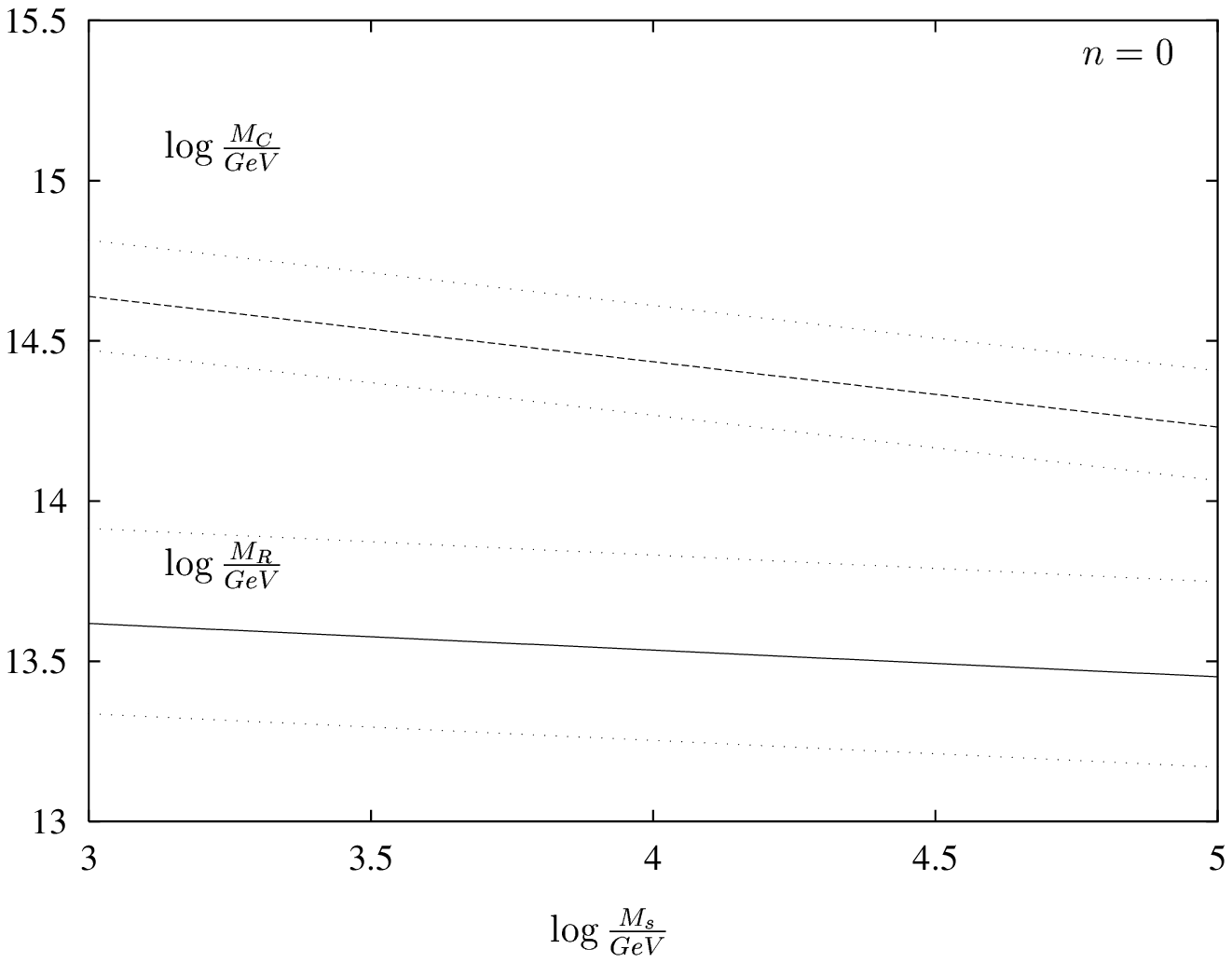,height=7cm}
\psfig{figure=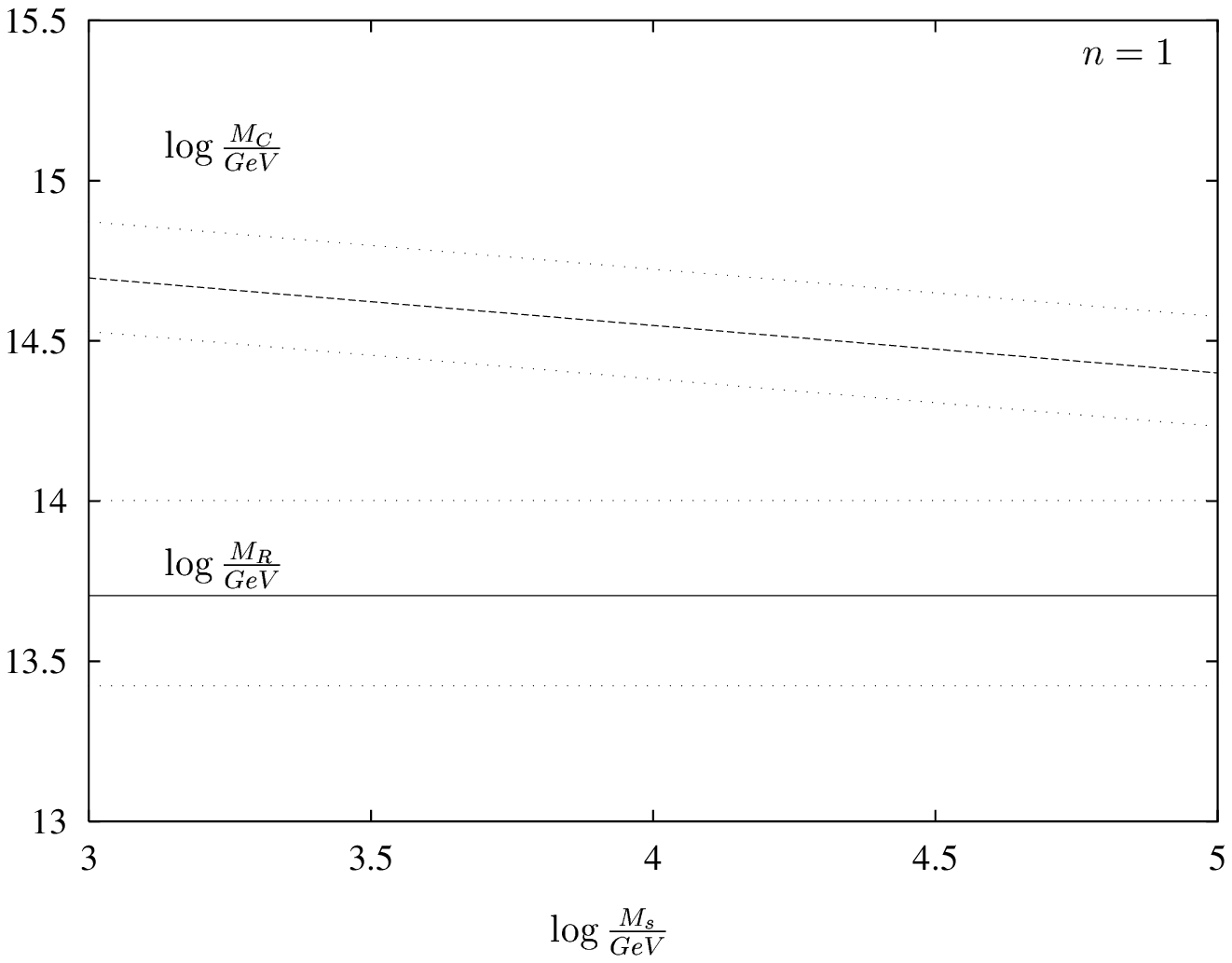,height=7cm}}
\caption{$M_C$ and $M_R$ vs. $M_s$ for Chain A, when $M_X$ is fixed at 
its lowest allowed value ($log(M_X/GeV)=15.5$) and $\epsilon = 1$, for
$n=0$ and $1$. Dotted lines are the same plots for lowest and highest
values of $\alpha_3$}.
\end{figure}


\vskip 1cm

\begin{references}

\bibitem{dva} T.~Kibble, G.~Lazarides and Q.~Shafi,
Phys. Rev. {\bf D26}, 435 (1982); D.~Chang, R.N.~Mohapatra
and M.K.~Parida, Phys. Rev. Lett. {\bf 52}, 1072 (1984).

\bibitem{seesaw} M.~Gell-Mann, P.~Ramond and R.~Slansky, 
in {\it Supergravity}, eds. P.~van~Niewenhuizen and D.Z.~
Freedman (North Holland 1979); T.~Yanagida, in Proceedings 
of {\it Workshop on Unified Theory and Baryon number in the 
Universe}, eds. O.~Sawada and A. Sugamoto (KEK 1979); 
R.N.~Mohapatra and G.~Senjanovi{\'c}, Phys. Rev. Lett. 
{\bf 44}, 912 (1980).  

\bibitem{ms81} 
R.N.~Mohapatra and G.~Senjanovi\'c, Phys. Rev. {\bf D23},165 (1981); 
M.~Magg and Ch.~Wetterich, Phys. Lett. {\bf B94} 61, (1980).

\bibitem{mohapatra99} R.N.~Mohapatra, hep-ph/9911272.

\bibitem{amrs99} C.S.~Aulakh, A.~Melfo, A.~Ra\v{s}in and 
G.~Senjanovi\'c, Phys. Lett. {\bf B459}, 557 (1999), hep-ph/9902409. 

\bibitem{abmrs99} C.S.~Aulakh, B.~Bajc, A.~Melfo, A.~Ra\v{s}in and
G.~Senjanovi\'c, Phys. Lett. {\bf B460}, 325 (1999), hep-ph/9904352.
                                              
\bibitem{ps74} J.C.~Pati and A.~Salam, Phys. Rev.{\bf D10}, 275 (1974).

\bibitem{lr}R.N.~Mohapatra and J.C.~Pati, Phys. Rev {\bf D11},
566 and 2558 (1975); G.~Senjanovi\'{c} and R.N.~Mohapatra, 
{\it ibid} {\bf D12}, 1502 (1975). For details see 
G.~Senjanovi\'c, Nucl. Phys. {\bf B153}, 334 (1979).

\bibitem{so10}
K.S.~Babu and S.M.~Barr, Phys. Rev. {\bf D51}, 2463 (1995),
hep-ph/9409285;\\
K.S.~Babu and R.N.~Mohapatra, Phys. Rev. Lett. {\bf 74}, 2418 (1995),
hep-ph/9410326;\\
G.~Dvali and S.~Pokorski, Phys. Lett. {\bf B379}, 126 (1996),
hep-ph/9601358;\\
S.M.~Barr and S.~Raby, Phys. Rev. Lett. {\bf 79}, 4748 (1997),
hep-ph/9705366;\\
Z.~Chacko and R.N.~Mohapatra, Phys. Rev. {\bf D59}, 011702 (1999),
hep-ph/9808458.

\bibitem{aulamoha} C.S.~Aulakh and R.N.~Mohapatra,
Phys. Rev. {\bf D28}, 217 (1983).

\bibitem{leem94}
D.-G.~Lee and R.N.~Mohapatra, Phys. Rev. {\bf D51}, 1353 (1995), 
hep-ph/9406328.

\bibitem{sato}
J.~Sato, Phys. Rev. {\bf D53}, 3884 (1996), hep-ph/9508269.

\bibitem{bdm98} K.S.~Babu, B.~Dutta and R.N.~Mohapatra,
Phys. Rev. {\bf D60}, 095004 (1999), hep-ph/9812421.

\bibitem{amrs98} C.S.~Aulakh, A.~Melfo, A.~Ra\v{s}in 
and G.~Senjanovi\'c, Phys. Rev {\bf D58}, 115007 (1998), 
hep-ph/9712551.

\bibitem{am82}
C.S.~Aulakh and R.N.~Mohapatra,
\newblock Phys. Lett. {\bf 119B}, 136 (1982).  

\bibitem{cein99} J.A.~Casas, J.R.~Espinosa, A.~Ibarra and I.~Navarro,
hep-ph/9905381.

\bibitem{dw83} S. Dimopoulos and F. Wilczek, in {\it The Unity of 
the Fundamental Forces}, ed. A. Zichichi, Plenum Press, New York, 1983.

\bibitem{babu94} K.S. Babu and S.M. Barr, Phys. Rev. {\bf D50}, 3529
(1994), hep-ph/9402291. 

\bibitem{hmy91} N.~Sakai and T.~Yanagida, Nucl. Phys. {\bf B197}, 
533 (1982); S.~Weinberg, Phys. Rev. {\bf D26}, 287 (1982); 
J.~Hisano, H.~Murayama and T.~Yanagida, Nucl. Phys. {\bf 402B}, 
46 (1993), hep-ph/9207279.

\bibitem{babu} A.~Ra\v{s}in, UC Berkeley Ph.D. Thesis, LBL-35586 
(1994); A.~Antaramian, UC Berkeley Ph.D. Thesis, LBL-36819 (1995); 
H.~Murayama and D.~Kaplan, Phys. Lett. {\bf 336B}, 221 (1994), 
hep-ph/9406423; V.~Ben-Hamo and Y.~Nir, Phys. Lett. {\bf 339B}, 
77 (1994), hep-ph/9408315; K.S.~Babu, J.~Pati and F.~Wilczek, 
Nucl. Phys. {\bf B566}, 33 (2000), hep-ph/9812538.

\bibitem{wilczek} K.S.~Babu, J.~Pati and F.~Wilczek, 
Phys. Lett. {\bf 423B}, 337 (1998), hep-ph/9712307.

\bibitem{ams98} C.S.~Aulakh, A.~Melfo and G.~Senjanovi\'c,
Phys. Rev. {\bf D57}, 4174 (1998), hep-ph/9907256.

\end{references}
\end{document}